\documentclass[10pt,twocolumn,amsmath,amssymb,superscriptaddress,footinbib]{revtex4-1}
\usepackage{amsmath}
\usepackage{graphicx}
\usepackage{xcolor}

\begin{document}

\title{\textit{In-situ} Raman study of laser-induced graphene oxidation}

\author{Felix Herziger}
\email{fhz@physik.tu-berlin.de}
\affiliation{Institut f\"ur Festk\"orperphysik, Technische Universit\"at Berlin, Hardenbergstrasse 36, 10623 Berlin, Germany}

\author{Rasim Mirzayev}
\affiliation{Institut f\"ur Festk\"orperphysik, Technische Universit\"at Berlin, Hardenbergstrasse 36, 10623 Berlin, Germany}
\affiliation{University of Vienna, Faculty of Physics, Boltzmanngasse 5, 1090 Vienna, Austria}

\author{Emanuele Poliani}
\affiliation{Institut f\"ur Festk\"orperphysik, Technische Universit\"at Berlin, Hardenbergstrasse 36, 10623 Berlin, Germany}

\author{Janina Maultzsch}
\affiliation{Institut f\"ur Festk\"orperphysik, Technische Universit\"at Berlin, Hardenbergstrasse 36, 10623 Berlin, Germany}

\begin{abstract}
We present \textit{in-situ} Raman measurements of laser-induced oxidation in exfoliated single-layer graphene. By using high-power laser irradiation, we can selectively and in a controlled way initiate the oxidation process and investigate its evolution over time. Our results show that the laser-induced oxidation process is divided into two separate stages, namely tensile strain due to heating and subsequent $p$-type doping due to oxygen binding. We discuss the temporal evolution of the $D/G$-mode ratio during oxidation and explain the unexpected steady decrease of the defect-induced $D$ mode at long irradiation times. Our results provide a deeper understanding of the oxidation process in single-layer graphene and demonstrate the possibility of sub-$\mu$m patterning of graphene by an optical method.
\end{abstract}

\maketitle%

\maketitle

\section{Introduction}
Graphene and related two-dimensional materials, such as transition metal dichalcogenides, have experienced increasing scientific interest during the last decade \cite{RevModPhys.81.109,10.1126/science.1158877,10.1038/nnano.2012.193}. Due to their unique properties, these materials are likely candidates for future applications in sensors, as well as nano- and opto-electronic devices \cite{10.1038/nnano.2014.215,10.1038/nnano.2014.207}. However, despite graphene's extraordinary high charge-carrier mobility \cite{10.1126/science.1102896}, the lack of an intrinsic bandgap prevents this material from integration in transistors and logic devices. A possible route to overcome this limitation is the precise and controllable modification of graphene's electronic properties using, for instance, oxidation or hydrogenation \cite{10.1126/science.1167130,10.1021/nl0808684,10.1021/ja9043906,10.1021/nn201226f,10.1021/nl1029607,10.1021/jp305823u,10.1038/srep04892,10.1021/nn503574p,10.1007/s12274-010-1015-3,10.1021/nn9012753}. Selective functionalization may also offer the possibility to design artificial graphene structures with tailored properties. In fact, it has been demonstrated that oxidation of graphene may open a bandgap \cite{10.1039/C4NR05207B}. However, the precise temporal evolution of graphene oxidation has not been reported so far and thus prevents a deeper understanding of this process.    

In this work, we present an \textit{in-situ} Raman study of the oxidation process in mechanically exfoliated single-layer graphene. We can selectively initiate the oxidation process by high-power laser irradiation and subsequently observe the temporal dependence of the $G$ and $2D$ modes. We observe a sharp increase of the $D/G$-mode ratio shortly after initiating the oxidation process, followed by a saturation of this ratio and a steady decrease afterwards. In fact, the $D$-mode reduces in intensity nearly down to its initial value prior to oxidation, \textit{i.e.}, it is almost absent in the Raman spectrum after $t > 1000$\,s . Simultaneously, we observe a strong luminescent background that behaves likewise. By correlating the measured $G$- and $2D$-mode positions, we provide deeper insights into the different stages of the oxidation process and are able to explain the observed temporal dependence of $D/G$-mode ratio. Finally, we discuss AFM measurements of the irradiated single-layer graphene and demonstrate sub-$\mu$m patterning of graphene.

\section{Experimental details}
Single-layer graphene samples were prepared by micro-mechanical exfoliation of natural graphite crystals onto silicon substrates with an 100\,nm thick oxide layer. Graphene samples were searched with an optical microscope and their layer number was determined using the optical contrast and layer-number dependent Raman modes \cite{10.1021/nn800307s,PhysRevB.85.235447,10.1021/nl302450s}. Raman measurements were carried out using a Horiba HR800 spectrometer equipped with a Nd:YAG laser with 532\,nm wavelength. The \textit{in situ} measurements were performed in ambient conditions with a time resolution of 1\,s over a period of more than 2000\,s. We used a 600\,lines/mm grating in order to record all important Raman modes within the same spectral window at each time frame. The laser power was chosen to be 40\,mW on the sample. Raman maps with low laser power after irradiation were obtained by a motorized \textit{xyz} stage with a minimum step size of 250\,nm and using a laser power of less than 1\,mW in order to avoid sample heating or additional structure modifications of the graphene layer.

The laser spot diameter (full width at half maximum - FWHM) in our studies was estimated from a simple, straight-forward measurement. We performed a linescan across the edge of a thick graphite flake on a SiO$_2$/Si substrate and recorded the intensity of the silicon Raman signal. The decrease of the silicon signal at the graphite edge gives a good estimate of the laser spot size. The spatially varying silicon signal was fitted by a cumulative distribution function, \textit{i.e.}, the convolution of a Gaussian curve and a Heaviside function. At 532\,nm laser wavelength with a power of 40\,mW and using a 100x objective (NA = 0.90), we determine a FWHM of the laser spot size of approximately 750\,nm.

Atomic force microscopy (AFM) images were recorded in true non-contact mode using a Park Systems XE 100 AFM. All measurements were performed in ambient conditions and at room temperature. AFM images were processed using the WSxM software \cite{10.1063/1.2432410}.

\begin{figure}[t]%
\centering
\includegraphics{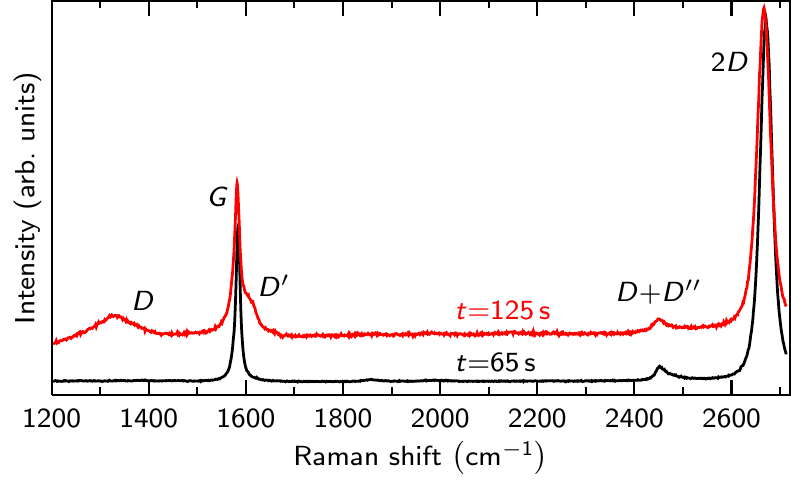}%
\caption{Raman spectra of exfoliated single-layer graphene under high-power laser irradiation at different times as given next to the spectra. Spectra are not scaled nor vertically offset. \label{fig1}}%
\end{figure}

\section{Results and discussion}
Figure~\ref{fig1} presents Raman spectra measured at different times during laser irradiation with 40\,mW laser power. As can be seen, the spectrum at $t=65$\,s only exhibits the well-known Raman modes in exfoliated single-layer graphene, \textit{i.e.}, the first-order $G$ mode, as well as the double-resonant $D+D''$ (iTOLA) and $2D$ modes. In contrast, the spectrum at $t=125$\,s exhibits a strong increase in its background intensity (note that the spectra are not scaled nor vertically offset). Furthermore, we observe the presence of the defect-related double-resonant $D$ and $D'$ Raman modes. In the following, we will discuss the different effects that are observed during laser irradiation in ambient conditions. 

\begin{figure*}[t]%
\centering
\includegraphics{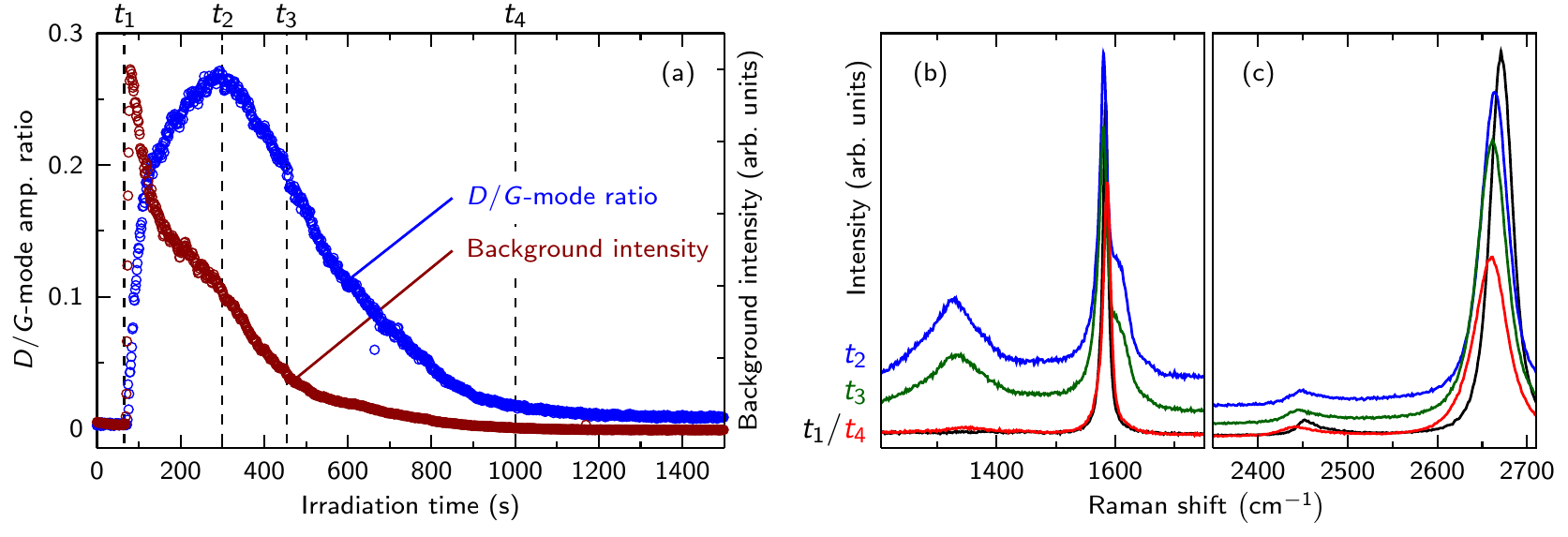}%
\caption{(a) Evolution of the $D$/$G$-mode amplitude ratio (blue circles) and the background intensity (red circles) in single-layer graphene with increasing irradiation time using high-power laser irradiation. (b),(c) Raman spectra in the $G$- and $2D$-mode spectral range for different irradiation times $t_i$ as defined in (a). Spectra are not scaled in intensity nor vertically offset. \label{fig2}}%
\end{figure*}

In Figure~\ref{fig2}\,(a), we present the evolution of the $D$/$G$-mode amplitude ratio in single-layer graphene with increasing time under high-power laser irradiation. As can be seen, until $t_1 = 65$\,s this ratio does not change. Afterwards, we observe an abrupt increase to a maximum value of approximately 0.27, followed by a nearly linear decrease with increasing irradiation time. For irradiation times $t>t_4 \approx 1000$\,s, the $D/G$-mode ratio is nearly constant and approaches the initial value of the exfoliated single-layer graphene. The corresponding spectra for each time $t_i$ from Fig.~\ref{fig2}\,(a) are given in Figs.~\ref{fig2}\,(b) and (c) for the $G$- and $2D$-mode spectral range, respectively. The spectra are not scaled in intensity nor vertically offset. The black spectrum at $t_1$ and the blue spectrum at $t_2$ exhibit a huge difference in their background signal, as could be already seen from Fig.~\ref{fig1}. Furthermore, we observe the appearance of intense defect-related Raman modes, \textit{i.e.}, the double-resonant $D$ and $D'$ modes at $\sim 1345$\,cm$^{-1}$ and $\sim 1615$\,cm$^{-1}$, respectively. With increasing irradiation time, both the background intensity and the defect-related Raman modes decrease in intensity. For irradiation times larger than $t_4$, the Raman spectrum of the graphene layer inside the laser spot nearly resembles the initial Raman spectrum at $t_0$, \textit{i.e.}, the $D$/$G$-mode ratio drops to values of less than 0.01. Furthermore, also the background intensity has approached the initial value. The main difference between the spectra at $t_0$ and $t_4$ is a significantly reduced $2D$/$G$-mode ratio of approximately 1.5 at $t_4$ and a downshifted $2D$-mode position, which will be discussed below [compare Fig.~\ref{fig2}\,(c)]. The broadening and downshift of the $D+D''$ mode follows the evolution of the $2D$ mode.

In Fig.~\ref{fig2}\,(a), we also plot the background intensity as a function of the irradiation time. Qualitatively, the same behavior as for the $D$/$G$-mode ratio is observed, \textit{i.e.}, a first sharp increase in intensity followed by a continuous decrease back to the initial level. However, the background intensity rises and drops more abruptly than the $D/G$-mode ratio. Furthermore, the background intensity increases slightly before the increase of the $D/G$-mode ratio and starts to drop as the $D/G$ ratio is rising. This may indicate that the origin of these effects are two competing processes. We attribute the strong and spectrally broad background in our spectra to luminescence from recombination of thermalized electron-hole pairs \cite{PhysRevB.82.121408,10.1038/srep02315}. By starting laser irradiation of our single-layer graphene sample, we effectively start heating the graphene layer. Furthermore, we also start heating the silicon substrate underneath. Since silicon dioxide is an amorphous, wide-bandgap insulator with a low thermal conductivity of approximately 1\,W/(m$\cdot$K) \cite{PhysRevB.50.6077}, the graphene layer is shielded from the heat of the silicon substrate at short irradiation times. However, at a certain time $t$ the heat will reach the SiO$_2$ surface and additionally heat the graphene layer. This will introduce short-range distortions and buckling in the graphene layer \cite{10.1021/nl1029607}. Thus, the local curvature of the graphene layer will increase, which drastically enhances the hot-carrier emission efficiency \cite{10.1038/srep02315}. Therefore, the first sharp increase in background intensity is given by the time, when the heat from the silicon substrate reaches the SiO$_2$ surface and graphene straining and buckling begins. In the following, the background intensity drops, while the $D/G$-mode ratio increases, \textit{i.e.}, defects are created. The decreasing background can be understood from the fact that the contribution from carrier-defect scattering increases and thus reduces the number of excited charge carriers. Furthermore, the creation of defects leads to a reduction of the local curvature of the graphene layer and therefore lowers the hot-luminescence emission efficiency. For $t>700$\,s, the background intensity has dropped below 5\,\% of its maximum value.

\begin{figure}[t]%
\centering
\includegraphics{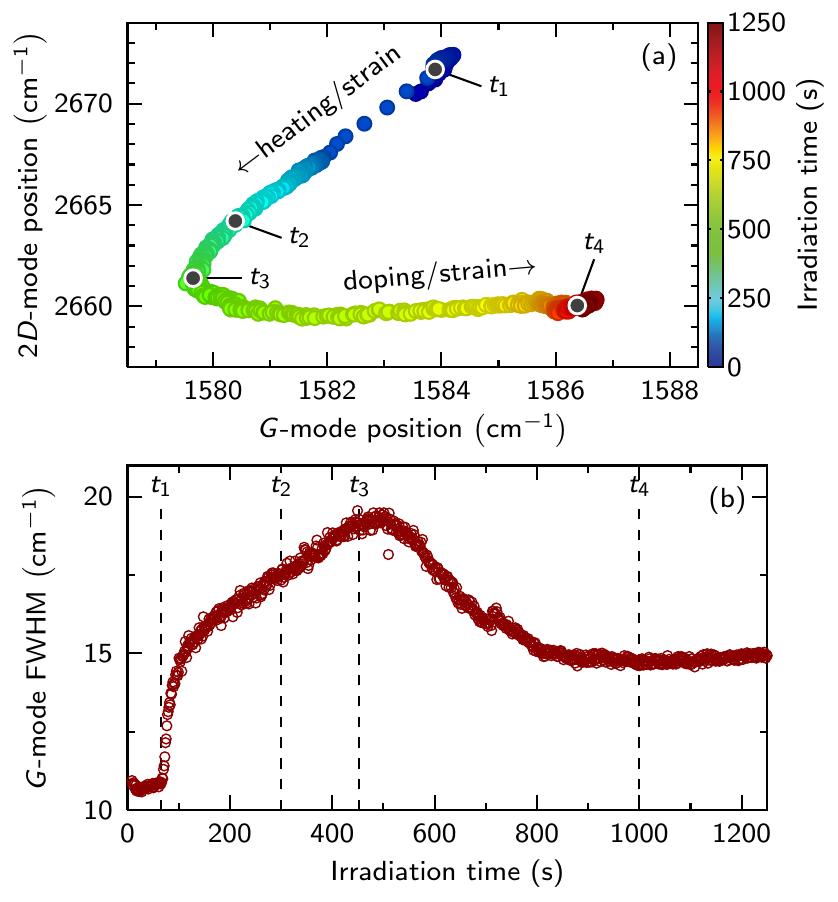}%
\caption{(a) Time evolution of the $2D$-mode position as a function of the $G$-mode position. Characteristic times from Fig.~\ref{fig2}(a) are marked with gray, filled circles and labeled with $t_i$. (b) FWHM of the $G$ mode as a function of irradiation time. The different $t_i$ are marked by vertical dashed lines. \label{fig3}}%
\end{figure}

\begin{figure*}[t]%
\centering
\includegraphics{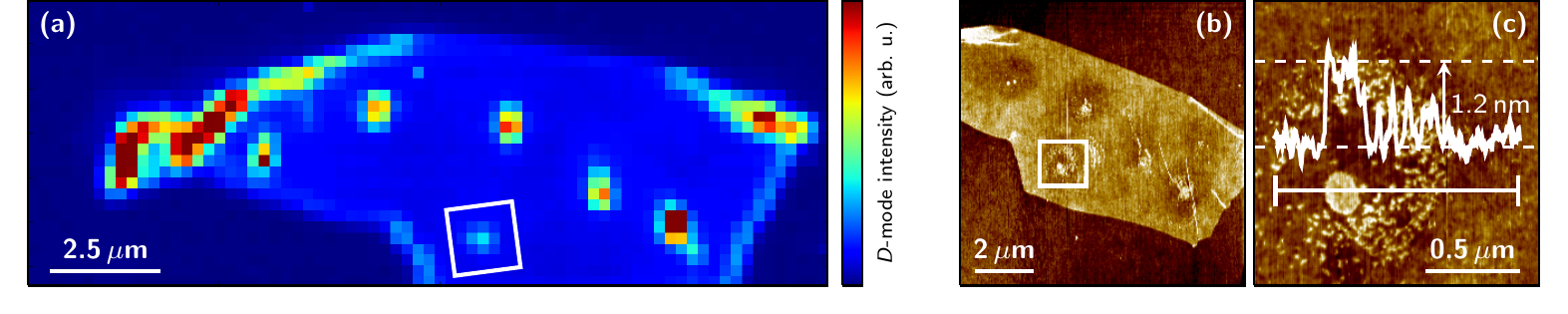}%
\caption{(a) Raman mapping of the $D$-mode intensity on a single-layer graphene flake that was treated by high-power laser irradiation. The irradiated regions can be identified as small circular spots. (b) Atomic-force microscopy image of the same flake as in (a). Again, the irradiated regions can be clearly identified as circular spots. (c) Enlarged view of the region that is marked in (a) and (b) by the white rectangle. The height profile has been recorded along the white, horizontal line. The height difference between graphene and the adsorbates is approximately 1.2\,nm, which corresponds to the height of single-layer graphene oxide. \label{fig4}}%
\end{figure*}

We will now turn to a discussion of the $G$- and $2D$-mode positions. Figure~\ref{fig3}\,(a) plots the peak positions of both Raman modes for all data points with $t\leq1250$; the time order of the data is indicated by the different colors of the data points. For times larger than $t_4$, we observe only minor peak shifts that can be attributed to further increasing tensile strain (not shown). As can be seen, for irradiation times between $t_1$ and $t_3$ both the $G$- and $2D$-mode frequencies decrease linearly with a slope of approximately $\Delta\omega(2D)/\Delta\omega(G) =2.2$. A slope with this value is commonly identified with uniaxial or biaxial tensile strain \cite{PhysRevB.79.205433,PhysRevB.80.205410,10.1038/ncomms2022,PhysRevApplied.2.054008}, however, biaxial strain is more likely to be present in our experiment. Using a strain-induced $G$-mode shift rate of $-50\,\text{cm}^{-1}/\%$ \cite{PhysRevApplied.2.054008}, we calculate a strain difference of $\Delta\epsilon \approx 0.1\,\%$ between $t_1$ and $t_3$. For irradiation times larger than $t_3$, the $G$ mode upshifts again, whereas the $2D$ mode shows a nearly constant peak position with only slight increase. The slope of these data points exhibits a value of approximately $\Delta\omega(2D)/\Delta\omega(G) =0.1$. Thus, this region cannot be identified with purely $n$-type (slope of $-0.1$) or purely $p$-type doping (slope of 0.4) \cite{10.1088/1367-2630/15/11/113006}. In fact, it is very reasonable that the tensile strain further increases from $t_3$ to $t_4$. Thus, the observed evolution of the $G$- and $2D$-mode positions is a superposition of tensile strain and doping effects. The further increasing strain will reduce the slope of the data points between $t_3$ and $t_4$, indicating that the observed correlation of the $G$- and $2D$-mode frequencies in this time interval is due to $p$-type doping. Since the experiments were performed in ambient conditions, functionalization by oxygen is very likely to occur. This would lead to $p$-type doping \cite{10.1021/nl1029607}, in accordance with our data. From the $G$-mode shift, we estimate a difference in carrier concentration between $t_3$ and $t_4$ of $\Delta n = 6\times 10^{12}\,\text{cm}^{-2}$ \cite{10.1038/ncomms2022}. The effect of doping can be also seen in the $G$-mode FWHM which is shown in Fig.~\ref{fig3}\,(b). After a first increase of the FWHM caused by the creation of defects, we observe a steady decrease of the FWHM between $t_3$ and $t_4$ due to doping. The linear decrease of the Raman mode positions between $t_1$ and $t_3$ due to tensile strain can be attributed to laser-induced heating of the graphene flake and the silicon substrate underneath \cite{10.1021/nl071033g,10.1021/nl201488g,PhysRevB.91.075426}. From the observed shift of the $G$ mode and the thermal expansion coefficient of graphene on SiO$_2$/Si of $-0.016\,\text{cm}^{-1}/^{\circ}C$ \cite{10.1021/nl071033g}, we estimate a temperature increase of approximately 250\,K inside the laser spot during irradiation.

As discussed in Ref.~\cite{10.1021/nl1029607}, thermal annealing of graphene supported on SiO$_2$/Si substrates introduces short-range distortions to the graphene lattice and therefore facilitates oxygen binding. Oxygen functionalization is further catalyzed by the likely presence of a partial water layer between graphene and the substrate, as well as a water adlayer on top of graphene \cite{10.1021/jp305823u}. Following Refs. \cite{10.1021/nl1029607} and \cite{10.1021/jp305823u}, the initially observed tensile strain results from a conformation of graphene to the SiO$_2$/Si substrate, leading to the creation of ripples. These surface ripples reduce the activation energy for oxygen binding \cite{10.1021/jp305823u}. Therefore, oxidation does not occur directly, but apparently needs a certain amount of strain to weaken the bonds and increase chemical reactivity \cite{10.1021/nl1029607}. Both effects can be clearly seen in Figure~\ref{fig3}\,(a).

Finally, we want to explain the unexpected decrease of the $D/G$-mode ratio with increasing irradiation time. In principle, one would expect increasing defect-related Raman modes due to adsorption of molecules from the air and the progressing oxygen binding to the basal plane of graphene. However, we identify two processes that antagonize this expectation. First, laser-induced annealing inside the laser spot and, second, the doping dependence of the $D/G$-mode ratio. It has been reported by many different works that annealing of graphene and/or carbon nanotubes leads to a reduction of physisorbed adsorbates and thus a decrease of the $D$-mode intensity \cite{10.1021/jp305823u,10.1021/nn800031m,10.1016/j.carbon.2011.11.010}. Since temperatures inside our laser spot reach values of approximately 550\,K, annealing seems likely to occur. Moreover, it has been experimentally demonstrated by Bruna \textit{et al.} \cite{10.1021/nn502676g} and Froehlicher \textit{et al.} \cite{PhysRevB.91.205413} that the $D/G$-mode ratio depends strongly on the doping level in single-layer graphene. This effect can be understood from an increased electronic broadening in the $D$-mode double-resonance Raman process, resulting in a decrease in intensity \cite{PhysRevB.80.165413,PhysRevB.84.035433}. Moreover, the $G$-mode intensity increases with either $n$- or $p$-type doping \cite{10.1021/nn1010914,10.1038/nature09866}. Both effects lead to a reduction of the measured $D/G$-mode ratio in doped graphene. As we have undoubtedly demonstrated in Figure~\ref{fig3}, we observe an increasing doping level in our graphene layer with increasing irradiation time due to chemisorbed oxygen. Thus, a doping-related decrease of the $D/G$-mode ratio is reasonable. In total, these two effects, \textit{i.e.}, annealing of physisorbed absorbates and doping by chemisorbed oxygen, lead to a reduction of the $D/G$-mode ratio with increasing irradiation time although oxidation continues and probably creates further defects.

In Figure~\ref{fig4}\,(a), (b), and (c), we present a Raman map (1\,mW laser power) of the $D$-mode intensity and AFM images of the same graphene flake that has been irradiated at different spatial positions. The Raman map is composed of 1898 individual Raman spectra with a point-to-point distance of 250\,nm. Inside the graphene flake, we can identify several regions that exhibit a significantly increased $D$-mode intensity compared to the surrounding regions. These regions correlate with the spatial positions, where we oxidized the graphene layer by high-power laser irradiation. The increased $D$-mode intensities at the left edge and at the top right corner correspond to regions of folded graphene [compare Fig.~\ref{fig4}\,(b)] and are not related to laser irradiation. In Figure~\ref{fig4}\,(b) we present an AFM image of the same graphene flake. We observe a close correspondence between the regions with high $D$-mode intensity and the structurally modified regions in the AFM image, \textit{i.e.}, the laser-modified regions can be again identified as circular regions. The diameter of these regions is approximately 970\,nm and in reasonable agreement with our laser spot size. Inside these regions, we observe small spots of laser-deposited material that show a drastically increased height compared to the surrounding area [see Figure~\ref{fig4}\,(c)]. The height of the laser-deposited material inside the laser spot is approximately 1.2\,nm, which nicely coincides with the height of graphene oxide reported in literature \cite{10.1038/nnano.2009.58}. This gives further evidence to laser-induced oxidation of our graphene samples on the sub-$\mu$m scale. Thus, by scanning the laser spot across the graphene flake, spatially controlled functionalization of graphene can be achieved. In principle, arbitrary structures can be realized, as demonstrated in similar experiments \cite{10.1039/C4NR05207B,10.1038/srep04892,10.1021/nn503574p,10.1021/nn201226f}.

\section{Conclusion}
In summary, we demonstrated an \textit{in-situ} analysis of the oxidation process in single-layer graphene. We showed that high-power laser irradiation in ambient conditions can be separated into two different stages; tensile strain due to laser-induced heating and subsequent $p$-type doping due to oxidation. The observed temporal decrease of the $D/G$-mode ratio with increasing irradiation time can be explained with laser-induced annealing and the doping dependence of the double-resonant $D$-mode scattering process. Our results provide a deeper understanding of basal-plane oxidation in graphene and demonstrate the possibility of tailoring graphene's properties selectively at the sub-$\mu$m scale using a fully optical method.

\begin{acknowledgements}
This work was supported by the European Research Council (ERC) under grant no. 259286 and by the Deutsche Forschungsgemeinschaft (DFG) under grant number MA 4079/7-2. 
\end{acknowledgements}

\bibliographystyle{apsrev4-1}

\end{document}